\documentclass[reprint, amsmath, amssymb, aps, superscriptaddress, nofootinbib, preprintnumbers]{revtex4-2}

\usepackage{graphicx} 
\usepackage{epstopdf}
\usepackage{bm}
\usepackage[colorlinks=true,linktocpage=true,linkcolor=blue,citecolor=blue,urlcolor=blue]{hyperref}
\usepackage{textcomp}
\usepackage{color}
\usepackage{xcolor}
\usepackage{tabularx}
\usepackage{tabu}
\usepackage{amsmath,amssymb,mathtools}
\usepackage{feynmp-auto}
\usepackage[normalem]{ulem}

\begin{document}

\preprint{IFT-UAM/CSIC-25-86}

\title{Complete NLO SMEFT Electroweak Corrections to Higgs Decays }

\def\BNL{High Energy Theory Group, Physics Department, 
    Brookhaven National Laboratory, Upton, NY 11973, USA}
    
\def\FSU{Physics Department, Florida State University, Tallahassee, FL 32306-4350, USA}

\def\IFT{Departamento de F\'{i}sica Te\'{o}rica and Instituto de F\'{i}sica Te\'{o}rica UAM/CSIC, Universidad Aut\'{o}noma de Madrid, Cantoblanco, 28049, Madrid, Spain}

\def\YITP{C. N. Yang Institute for Theoretical Physics, Stony Brook University, Stony Brook, NY, 11794, USA}

\author{Luigi Bellafronte}
\email{lbellafronte@fsu.edu}
\affiliation{\FSU}

\author{Sally Dawson}
\email{dawson@bnl.gov}
\affiliation{\BNL}

\author{Clara Del Pio}
\email{cdelpio@bnl.gov}
\affiliation{\BNL}

\author{Matthew Forslund}
\email{mforslund@princeton.edu}
\affiliation{\BNL}
\affiliation{\YITP}

\author{Pier Paolo Giardino}
\email{pier.giardino@uam.es}
\affiliation{\IFT}
\begin{abstract}
\noindent
Precise predictions for Higgs decays are a crucial ingredient of the search for beyond the Standard Model (BSM) physics and the Standard Model Effective Field Theory (SMEFT) is a valuable tool for quantifying deviations from the Standard Model (SM).  We present the complete set of predictions for the 2- and 3- body Higgs decays at next-to-leading order (NLO), considering QCD and electroweak corrections and including all contributions from the dimension-6 SMEFT operators and with an arbitrary flavor structure.  Including the NLO SMEFT results for Higgs decays greatly increases the sensitivity to BSM physics of the $e^+e^-\rightarrow Zh$ process at FCC-ee, as compared with that obtained using only the total cross section.
\end{abstract}

\maketitle

\section{Introduction}

The study of Higgs properties is central to current and future collider experiments and  at the present time measurements of Higgs production and decay are in excellent agreement with theoretical predictions at the LHC.  Deviations from the Standard Model (SM) can be parameterized in terms of the Standard Model Effective Field Theory (SMEFT)~\cite{Brivio:2017vri}, in which the Lagrangian is expanded in terms of higher dimension operators whose effects are suppressed by powers of a large new physics scale, $\Lambda$. The SMEFT provides a consistent and model-independent framework to combine results from diverse data sets within global fits and to obtain limits on physics beyond the SM at high scales.  The global fits typically include LHC data,  $Z$ and $W$ pole observables, and di-boson production both at the LHC and LEP~\cite{deBlas:2025xhe,Celada:2024mcf,Ellis:2020unq,deBlas:2022ofj}.

Global fits in the context of the SMEFT often are performed using next-to-leading order (NLO) QCD calculations that can be automated~\cite{Degrande:2020evl,Brivio:2020onw}, while NLO electroweak (EW) corrections must be computed on a case by case basis.  The complete set of NLO EW SMEFT calculations needed for the global fits  do not exist, and here we present
an important step in this direction.  We compute the  NLO EW and QCD corrections for all 2- and 3- body  Higgs decays.  Using the narrow width approximation and previously computed NLO EW SMEFT results for $Z$ and $W$ decays~\cite{Dawson:2019clf, Bellafronte:2023amz}, the Higgs decay rates to 4-fermion final states that are accurate to  NLO in the dimension-6 SMEFT are obtained.
The impact of these corrections is illustrated for the Higgstrahlung process, $e^+e^-\rightarrow Z h$, at the future FCC-ee collider~\cite{Bernardi:2022hny,FCC:2025lpp}~\footnote{The sensitivity at the future CEPC will be similar, but we have not optimized for polarization, which could increase the sensitivity~\cite{CEPCPhysicsStudyGroup:2022uwl}}. The sensitivity from a measurement of the total cross section has been previously presented~\cite{Asteriadis:2024xuk,Asteriadis:2024xts} and here we demonstrate the impressive improvement in accuracy when the NLO SMEFT corrections for the Higgs decays are included using the narrow width approximation.

\section{SMEFT NLO Calculation}
\label{sec:smeft}

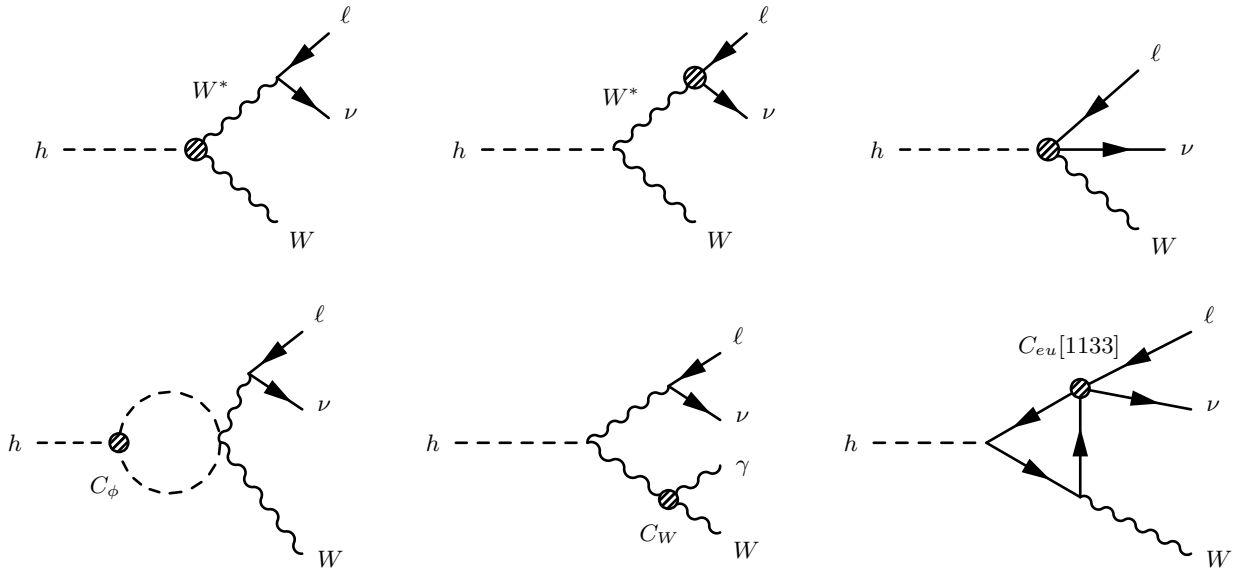
\begin{figure*}
    \centering
\begin{fmffile}{lo_ww}
  \begin{fmfgraph*}(100,120)
    \fmfstraight
    \fmfleft{i2,g,i1}
    \fmfright{o4,o3,o2,o1}
    \fmftop{top}
    \fmfbottom{bot}
    \fmf{dashes,tension=1.6}{g,h}
    \fmf{photon,tension=1.3,label=$W^*$,l.s=right}{v1,h}
    \fmf{photon,tension=1.3,l.s=right}{h,v2}
    \fmf{phantom,tension=1.0}{top,v1}
    \fmf{phantom,tension=1.0}{v2,bot}
    \fmfshift{16 down}{o1}
    \fmfshift{ 8 down}{o2}
    \fmfshift{ 8 up}{o3}
    \fmfshift{16 up}{o4}
    \fmf{fermion,tension=1.8}{o1,v1,o2}
    \fmf{phantom,tension=1.8}{o4,v2,o3}
    \fmfblob{0.08w}{h}
    \fmfv{l=$h$}{g}
    \fmfv{l=$\ell$}{o1}
    \fmfv{l=$\nu$}{o2}
    \fmfv{l=$W$}{v2}
  \end{fmfgraph*}
\end{fmffile} 
\hspace{1.5cm} 
\begin{fmffile}{lo_w}
  \begin{fmfgraph*}(100,120)
    \fmfstraight
    \fmfleft{i2,g,i1}
    \fmfright{o4,o3,o2,o1}
    \fmftop{top}
    \fmfbottom{bot}
    \fmf{dashes,tension=1.6}{g,h}
    \fmf{photon,tension=1.3,label=$W^*$,l.s=right}{v1,h}
    \fmf{photon,tension=1.3,l.s=right}{h,v2}
    \fmf{phantom,tension=1.0}{top,v1}
    \fmf{phantom,tension=1.0}{v2,bot}
    \fmfshift{16 down}{o1}
    \fmfshift{ 8 down}{o2}
    \fmfshift{ 8 up}{o3}
    \fmfshift{16 up}{o4}
    \fmf{fermion,tension=1.8}{o1,v1,o2}
    \fmf{phantom,tension=1.8}{o4,v2,o3}
    \fmfblob{0.08w}{v1}
    \fmfv{l=$h$}{g}
    \fmfv{l=$\ell$}{o1}
    \fmfv{l=$\nu$}{o2}
    \fmfv{l=$W$}{v2}
  \end{fmfgraph*}
\end{fmffile}
\hspace{1.5cm}
\begin{fmffile}{lo_4pt}
  \begin{fmfgraph*}(100,120)
    \fmfleft{i0}
    \fmfright{o3,o2,o1}
    \fmftop{top}
    \fmfbottom{bot}
    \fmf{dashes,tension=2}{i0,v1}
    \fmfshift{30 down}{o1}
    \fmfshift{30 up}{o3}
    \fmf{fermion,tension=1}{o1,v1,o2}
    \fmf{photon}{v1,o3}
    \fmfblob{8}{v1}
    \fmfv{l=$h$}{i0}
    \fmfv{l=$\ell$}{o1}
    \fmfv{l=$\nu$}{o2}
    \fmfv{l=$W$}{o3}
  \end{fmfgraph*}
\end{fmffile}
\vspace{1cm}
\begin{fmffile}{nlo_cphi}
\begin{fmfgraph*}(100,100)
    \fmfstraight
    \fmfleft{i1,i2,i3}
    \fmfright{o4,o3,o2,o1}
    \fmfshift{8 down}{o1}
    \fmfshift{ 4 down}{o2}
    \fmfshift{ 4 up}{o3}
    \fmfshift{8 up}{o4}
    \fmftop{top}
    \fmfbottom{bot}
    \fmf{dashes,tension=5}{i2,v1} 
    \fmf{dashes,tension=2,left}{v1,v2,v1}
    \fmf{boson,tension=2}{v2,t1}
    \fmf{phantom,tension=2}{v2,t2}
    \fmf{fermion,tension=2}{t1,o2}
    \fmf{fermion,tension=2}{o1,t1}
    \fmf{boson,tension=2}{v2,o4}
    \fmf{phantom,tension=2}{v2,o1}
    \fmf{phantom,tension=1}{t2,o3}
    \fmf{phantom,tension=1}{t2,o4}
    \fmf{phantom,tension=2}{t1,top}
    \fmf{phantom,tension=2}{t2,bot}
    \fmfblob{7}{v1}
    \fmfv{l.d=14,l.a=-112,l=\color{black}{$C_{\phi}$}}{v1}
    \fmfv{l=$h$}{i2}
    \fmfv{l=$\ell$}{o1}
    \fmfv{l=$\nu$}{o2}
    \fmfv{l=$W$,l.a=-20}{o4}
  \end{fmfgraph*}
\end{fmffile}  
\hspace{1.5cm}
\begin{fmffile}{nlo_rad}
  \begin{fmfgraph*}(100,100)   
    \fmfstraight
    \fmfright{o4,o3,o2,o1}
    \fmfleft{i2,h,i1}
    \fmftop{top}
    \fmfbottom{bot}
    \fmf{dashes,tension=1.6}{h,t1}
    \fmf{boson,tension=1.3}{t1,t3}
    \fmf{boson,tension=1.3}{t1,v1}
    \fmf{phantom,tension=1}{v1,top}
    \fmf{phantom,tension=1}{t3,bot}
    \fmfshift{16 down}{o1}
    \fmfshift{ 8 down}{o2}
    \fmfshift{ 8 up}{o3}
    \fmfshift{16 up}{o4}
    \fmf{fermion,tension=1.8}{o1,v1,o2}
    \fmf{boson,tension=1.8}{o4,t3,o3}
    \fmfv{l=$\ell$}{o1}
    \fmfv{l=$\nu$}{o2}
    \fmfv{l=$\gamma$}{o3}
    \fmfv{l=$h$}{h}
    \fmfv{l=$W$}{o4}
    \fmfblob{7}{t3} 
    \fmfv{l.d=10,l.a=-110,l=\color{black}{$C_W$}}{t3}
  \end{fmfgraph*}
  \hspace{1.5cm}
\end{fmffile}
\begin{fmffile}{nlo_ceett}
  \begin{fmfgraph*}(120,100)
    \fmfstraight
    \fmfright{o4,o3,o2,o1}
    \fmfleft{i2,h,i1}
    \fmftop{top}
    \fmfbottom{bot}
    \fmf{dashes,tension=1}{h,t1}
    \fmf{fermion,tension=.6}{t1,t3,t2,t1}
    \fmf{phantom,tension=.8}{i1,t2}
    \fmf{phantom,tension=.8}{t3,i2}
    \fmfshift{8 down}{o1}
    \fmfshift{ 4 down}{o2}
    \fmfshift{ 4 up}{o3}
    \fmfshift{8 up}{o4}
    \fmf{fermion,tension=1}{o1,t2,o2}
    \fmf{phantom,tension=1}{o3,t3}
    \fmf{boson,tension=1}{o4,t3}
    \fmfblob{7}{t2}
    \fmfv{l.d=12,l.a=110,l=\color{black}{$C_{eu}[1133]$}}{t2}
    \fmfv{l=$h$}{h}
    \fmfv{l=$\ell$}{o1}
    \fmfv{l=$\nu$}{o2}
    \fmfv{l=$W$,l.a=-20}{o4}
  \end{fmfgraph*}
\end{fmffile}
\vspace{-0.5cm}
\caption{Top: diagrams contributing to $h\rightarrow W l \nu$ at LO in the dimension-6 SMEFT. Bottom: representative virtual and real emission diagrams from SMEFT dimension-6 operators that first arise at one-loop. The circles represent dimension-6 SMEFT insertions.}
\label{fig:feynlo}
\end{figure*}

The dimension-6 SMEFT Lagrangian is defined as an $SU(3)\times SU(2)\times U(1)$ gauge invariant expansion around the SM
\begin{equation}
L=L_{SM}+\Sigma_i{C_i\over\Lambda^2}O_i+...
\end{equation}
where the operators $O_i$ contain only SM fields, $\Lambda$ is the scale of new physics and all information about  BSM physics resides in the coefficient functions, $C_i$.  We use the Warsaw basis~\cite{Grzadkowski:2010es} to define the dimension-6 operators and make no flavor assumptions other than that the CKM matrix be diagonal.

We  compute the complete set of 2- and 3- body processes that are relevant for Higgs decays at NLO EW and NLO QCD in the dimension-6 SMEFT and truncate the SMEFT expansion at ${1/\Lambda^2}$. The processes that we include are

\begin{itemize}
\item $h\rightarrow f {\overline{f}}$
\item $h\rightarrow Z f {\overline {f}}$
\item
$h\rightarrow W f {\overline{f}}^\prime$
\item
$h\rightarrow gg$, $h\rightarrow \gamma\gamma$, $h\rightarrow \gamma Z$
\item $h\rightarrow Zgg$\, ,
\end{itemize}
where  $f=e,\mu,\tau, u, d, s, c, b$.
The NLO SMEFT corrections to $h\rightarrow f{\overline{f}}$ can be found in \cite{Cullen:2020zof,Cullen:2019nnr,Gauld:2016kuu}, and those for $h\rightarrow Z e^+e^-$ and $h\rightarrow Z \mu^+\mu^-$in \cite{Dawson:2024pft}. The Higgs decays to two gauge bosons, $\gamma\gamma$~\cite{Dawson:2018liq,Dedes:2018seb,Hartmann:2015aia}, $\gamma Z$~\cite{Dawson:2018pyl,Dedes:2019bew}, and $gg$~\cite{Martin:2023fad,Corbett:2021cil}  are known to one-loop accuracy.  The result for $h\rightarrow Zgg$ can be found from \cite{Rossia:2023hen}.

The NLO dimension-6 SMEFT calculations of $h\rightarrow W f {\overline{f}}^\prime$ and 
of the quark contribution to $h\rightarrow Z f{\overline{f}}$  are  new results of this work. As an example, we show in the first row of Fig.~\ref{fig:feynlo} the relevant LO diagrams for the calculation of the $h \to W \ell \nu$ process, while we present some sample NLO diagrams in the second row. In particular, at NLO we have many additional operators that contribute, such the Higgs tri-linear vertex and 4-fermion top quark operators, entering the virtual corrections (first and third diagram) as well as new contributions to the real photon emission, e.g. the triple gauge boson vertex (in the middle).  

We take as input parameters $G_\mu$, $M_W$, and $M_Z$ and use a hybrid scheme where the coefficient functions $C_i$ are renormalized in the ${\rm \overline{MS}}$ scheme~\cite{Jenkins:2013zja,Jenkins:2013wua,Alonso:2013hga} while SM parameters are defined on-shell. The SMEFT relation between  $G_\mu$  and the Higgs vev can be found in the appendix of Ref.~\cite{Dawson:2018pyl}. 
 We use dimensional regularization to treat infrared divergences in the virtual diagrams and dipole subtraction for the corrections from real photon and gluon emission~\cite{Denner:2019vbn}.

The complete set of NLO EW and QCD results for the dimension-6 Higgs decays have been implemented in a flexible Monte Carlo code, which gives as outputs the complete NLO SMEFT decay width for the Higgs boson, $\Gamma_{tot}$, and all branching ratios for the 2- and 3-body decays, along with differential distributions for the 3-body decays.  The code will be documented in a forthcoming paper.
A result that is consistent to  LO can be found by computing the widths for Higgs decays, $h\rightarrow (f_1 {\overline {f}_2})(f_3 {\overline {f}_4})$, exactly without using the narrow-width approximation and employing the complex-mass scheme to treat the off-shell $W/Z$ bosons in a gauge-invariant way.
An NLO accurate result for Higgs decays to the experimentally measured 4-fermion final states can be found consistently by using the narrow width approximation,
\begin{equation}
\Gamma_{NLO} \biggl(h\rightarrow (f_1 {\overline {f}_2})(f_3 {\overline {f}_4})\biggr)=
\Gamma\biggl(h\rightarrow V f_1 {\overline {f}_2}\biggr)\frac{\Gamma(V\rightarrow f_3 {\overline {f}_4})}
{\Gamma (V\rightarrow {\rm all})}
\label{eq:nwa}
\end{equation}
where $V=W,Z$.
The NLO EW and QCD results for $Z$ and $W$ decays are in \cite{Dawson:2019clf,Biekotter:2025nln} and the result of eq.~\ref{eq:nwa} must be expanded consistently to  ${\cal{O}}({1/\Lambda^2})$\cite{Dawson:2024pft}.

We present our results by combining Higgs production and decay in the narrow width approximation,
\begin{equation}
\sigma \biggl(e^+ e^- \rightarrow Z XX \biggr)= \sigma \biggl(e^+ e^- \rightarrow Z h \biggr)
\frac{\Gamma(h \rightarrow XX)}
{\Gamma (h\rightarrow {\rm all})} \, ,
\label{eq:nwa_1}
\end{equation}
where $XX$ stands for any of the channels listed in Sec.~\ref{sec:smeft}, either 2- or 3-body decays.

\section{Results}

As inputs, we take $m_H=125.1~\textrm{GeV}$, $m_t=172.5~\textrm{GeV}$, $m_b=4.92~\textrm{GeV}$, $M_W^{exp}=80.379~\textrm{GeV}$, and $M_Z^{exp}=91.1876~\textrm{GeV}$. 
The vector boson masses, $M_W$ and $M_Z$, are defined following \cite{Freitas:2023iyx, Bardin:1988xt}, as we adopt the complex mass scheme for the treatment of finite width effects in the $h\to \bar{f} f$ decays,\footnote{In the SMEFT, the real corrections to $h\to \bar{f}f$ include a diagram with a $Z$-boson propagator that can go on shell. We avoid the resulting divergence by consistently incorporating the $Z$ width through the complex mass scheme.}
\begin{align}
\label{eq:MZMZMZ}
    \begin{split}
    M_Z&=\frac{M_Z^{\exp}}{\sqrt{1+\left({\Gamma_Z^{\exp}}/{M_Z^{\exp}}\right)^2}} = 91.1535~{\rm GeV} \, , \\
    M_W&=\frac{M_W^{\exp}}{\sqrt{1+\left({\Gamma_W^{\exp}}/{M_W^{\exp}}\right)^2}} = 80.352~{\rm GeV}\, .
    \end{split}
\end{align}
Corrections to \eqref{eq:MZMZMZ} in the SMEFT are suppressed by terms of ${\cal{O}}\big(\frac{C M_Z\Gamma_Z}{\Lambda^2}\big)$ which we consistently neglect.
With these inputs, 
we compute the total Higgs width at NLO\footnote{Our tree level widths agree with those of \cite{Brivio:2019myy}.}
and all Higgs branching ratios for the processes listed in Sec.~\ref{sec:smeft} to NLO electroweak and QCD order in the dimension-6 SMEFT. In this letter we have fixed the renormalization scale at $\mu=m_H$.

We report here the expression
for the Higgs total width,
\begin{align}
&{\Gamma_{tot}(h\rightarrow {\textrm{all}})\over \Gamma_{tot}(h\rightarrow {\textrm{all}})_{SM}}=
1-0.0700 C_{\phi B}+0.126 C_{\phi \square} \\ \nonumber
&-0.0306 C_{\phi D}+1.62 C_{\phi G}-0.0576 C_{\phi W}+0.0583 C_{\phi WB} \\ \nonumber
&-0.0223 C_{dG}[33]-4.46 C_{d\phi}[33]+0.19 C_{dW}[33] \\ \nonumber
&-0.588 C_{e\phi}[33]-0.0103 C_{lequ}^{(1)}[3333]+0.071 C_{ll}[1111] \\ \nonumber
&-0.149 C_{\phi l}^{(3)}[11]-0.0118 C_{\phi ud}[33]-0.0333 C_{qu}^{(1)}[1331] \\ \nonumber
&-0.0444 C_{qu}^{(8)}[1331]+0.0988 C_{quqd}^{(1)}[3333] \\ \nonumber
&+0.0188 C_{quqd}^{(8)}[3333]-0.0355 C_{uG}[33] \\ \nonumber
&+0.00148 C_{\phi}
\end{align}
where we have adopted the $U(2)^5$ symmetric limit and neglected coefficients smaller than $10^{-2}$, with the exception of $C_\phi$, which is phenomenologically relevant, for the sake of compactness. The operators with the small coefficients are predominantly the 2- and 4-fermion operators not involving the top quark. The indices in square brackets are flavor indices, and we have taken $\Lambda = 1$ TeV.
Numerical values for   the coefficients of all dimension-6 operators that contribute, without the $U(2)^5$ flavor assumption, are given in the supplemental material.

At an $e^+e^-$ collider, we can perform a consistent study including NLO SMEFT effects in both production and decay. 
The dependence of precise measurements of the Higgstrahlung cross section on variations of the SMEFT coefficients at NLO was found in \cite{Asteriadis:2024xts,Asteriadis:2024xuk}. The sensitivity to SMEFT operators can be improved significantly by tagging the Higgs decays and incorporating the NLO results also for the decays.  The projected FCC-ee sensitivities for the various Higgs decay channels are given in Tab. \ref{tab:fcc}~\cite{Altmann:2025feg}.

\begin{table}[t!]
\begin{tabular}{|c|c|c|}
\hline
 & $\sqrt{s}=240~\textrm{GeV}$ & $\sqrt{s}=365~\textrm{GeV}$\\
 \hline\hline
 $b\overline{b}$ & 0.21 & 0.38\\
 \hline 
 $c\overline{c}$ & 1.6 & 2.9\\
 \hline 
 $s\overline{s}$ & 120 & 350\\
 \hline 
 $gg$ & 0.8 & 2.1\\
 \hline 
 $\tau^+\tau^-$ & 0.58 & 1.2\\
 \hline 
 $\mu^+\mu^-$ & 11 & 25\\
 \hline 
 $W W^*$ & 0.8 & 1.8\\
 \hline 
 $ZZ^*$ & 2.5 & 8.3\\
 \hline 
 $\gamma\gamma$ & 3.6 & 13\\
 \hline
 $Z\gamma$ & 11.8 & 22\\
 \hline
\end{tabular}
\caption{Uncertainty in $\%$ on Higgstrahlung measurements  including Higgs decays  at FCC-ee with $10.8$~ab$^{-1}$ ($3.12$~ab$^{-1}$) at $\sqrt{s}=240$ GeV ($\sqrt{s}=365$ GeV)~\cite{Altmann:2025feg}.}
\label{tab:fcc}
\end{table}

\begin{figure*}
	\centering
    \hspace{-25pt}
        \includegraphics[width=0.381\textwidth]{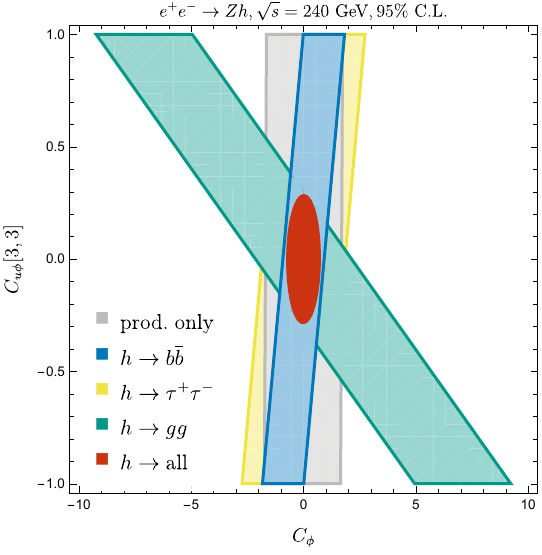} 
         \hspace{40pt}
        \includegraphics[width=0.38\textwidth]{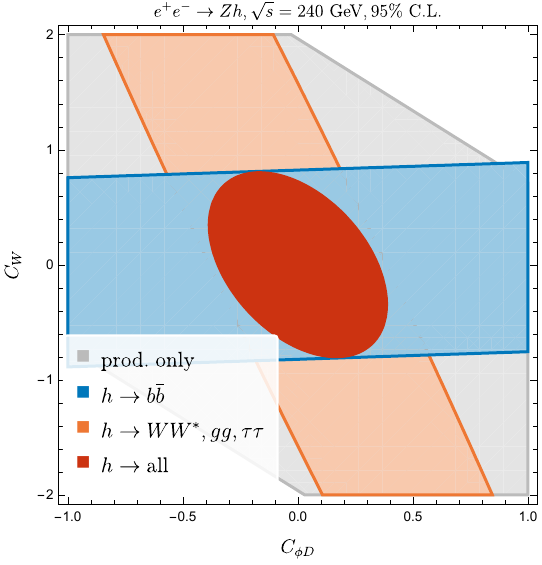}
	\caption{Left: $95\%$ confidence level sensitivity to $C_{u \phi}[3,3]$ and $C_\phi$ assuming $\Lambda=1$ TeV. The grey band assumes a $.5\%$ measurement of the total Higgstrahlung cross section, while the other bands assume the precision on the Higgs decays given in Tab. \ref{tab:fcc}. The red circle includes all measured Higgs decays.  Right: bounds on $C_{W}$ and $C_{\phi D}$.  }
	\label{fig:corr_240}
\end{figure*}

\begin{figure*}
	\centering  
    \hspace{-25pt}
        \includegraphics[width=0.38\textwidth]{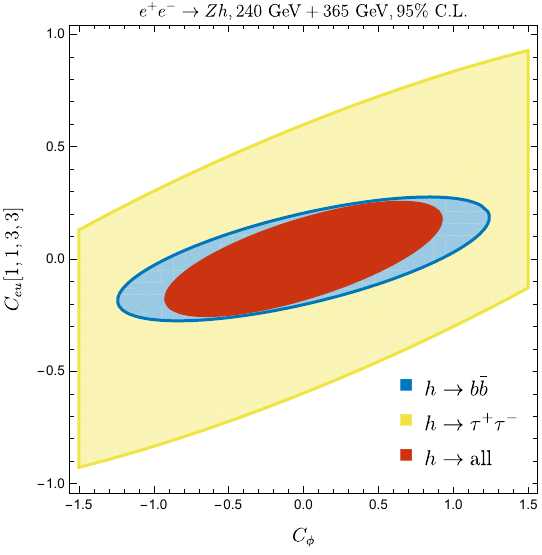}
         \hspace{40pt}
        \includegraphics[width=0.38\textwidth]{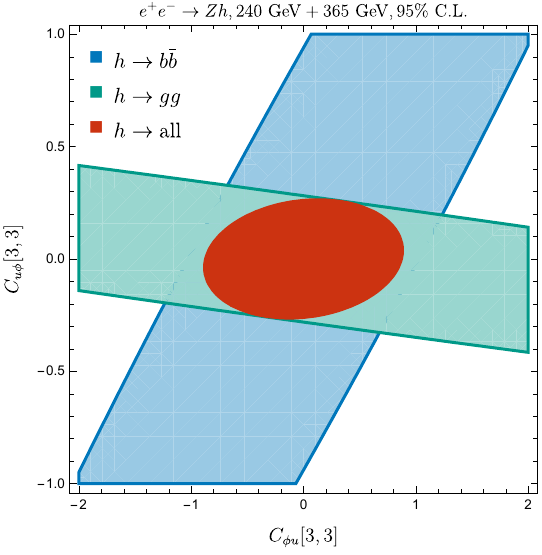
        }
	\caption{Left: $95\%$ confidence level limits on $C_{e u}[1,1,3,3]$ and $C_\phi$. Right: sensitivity to $C_{u \phi}[3,3]$ and $C_{\phi u}[3,3]$. We combine Higgstrahlung production at two energies, $\sqrt{s} = 240$ and $\sqrt{s} = 365$ GeV.
    }
	\label{fig:corr_240+365}
\end{figure*}

\begin{figure*}
	\centering  
    \hspace{-25pt}
        \includegraphics[width=0.38\textwidth]{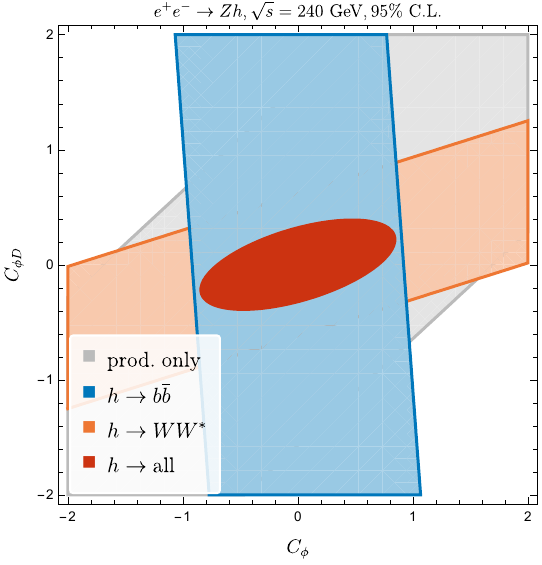}
         \hspace{40pt}
        \includegraphics[width=0.40\textwidth]{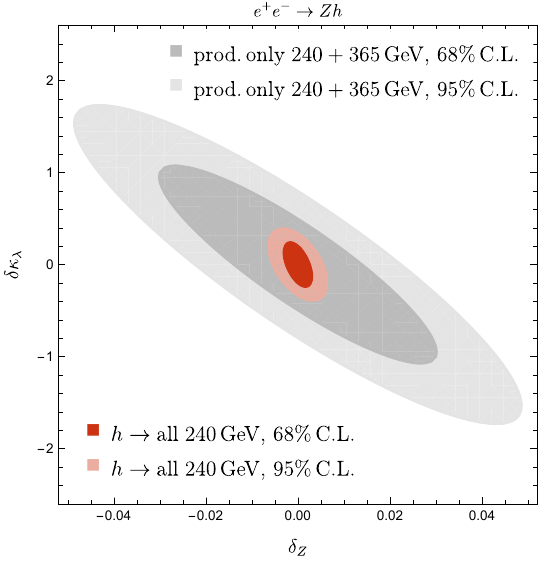}
	\caption{Bounds on coefficients contributing to $hZZ/hWW$ and tri-linear couplings. Left: SMEFT calculation in terms of $C_{\phi D}$ and $C_\phi$. Right: interpretation in the framework that makes use of the parameters $\delta_Z$ and $\delta \kappa_\lambda$ computed to linear order,  with $C_{\phi\square}$ set to $0$.} 
	\label{fig:kappa}
\end{figure*}

We present the correlations between constraints on different coefficient functions, obtained by combining production and decay, in Figs.~\ref{fig:corr_240}$-$\ref{fig:kappa}. For each plot, we set all but the displayed coefficients to $0$.
The left panel of Fig.~\ref{fig:corr_240} shows the sensitivity to $C_{u \phi}[3,3]$ and $C_\phi$. We consider the $e^+ e^- \to Zh $ production channel at $240$~GeV, assuming a $0.5\%$ accuracy on the FCC-ee cross section measurement for this process, and compare the sensitivity curve for production only (in grey) with the one obtained by taking into account also the Higgs decays (in red). The main contribution comes from the $b \bar{b}$ and $gg$ channels, which allow us to significantly constrain $C_\phi$ and $C_{u \phi}[3,3]$, respectively.
In the right panel one can see the limits on $C_W$ and $C_{\phi D}$. Adding the information on SMEFT effects from the Higgs decays, in particular the $h \to WW^*$contribution calculated here for the first time, offers a sensitive tool to constrain new
physics~\footnote{Since in~\cite{Altmann:2025feg} the uncertainty on $h\rightarrow W W^*$ is given considering only the hadronic decay channels of the $W$ boson, we use $h\rightarrow W W^*$ to denote the sum of the hadronic decays, i.e. $h\rightarrow W ud$, $h\rightarrow W cs$, and we sum over the charge of the $W$ bosons.}.

Fig.~\ref{fig:corr_240+365} shows the sensitivity curves obtained by combining Higgstrahlung production, followed by the Higgs decay, at two center-of-mass energies, $\sqrt{s} = 240$ GeV and $\sqrt{s} = 365$ GeV.  We assume uncertainties of $0.5\%$ and $1\%$ on the FCC-ee measurement of the $e^+ e^- \to Zh$ cross section at $\sqrt{s} = 240$ GeV and $\sqrt{s} = 365$ GeV, respectively. The reason for considering two energy runs to gain constraining power on $C_{eu}[1,1,3,3]$ and $C_{\phi u}[3,3]$ has been demonstrated in~\cite{Asteriadis:2024xts,Asteriadis:2024xuk}.
On the left-hand side, we present the correlation between limits on $C_{eu}[1,1,3,3]$ and $C_{\phi}$.
A key role is played by $h \to b \bar{b}$. The run at $\sqrt{s} =  240$~GeV is crucial to constrain $C_{\phi}$, as also shown in Fig.~\ref{fig:corr_240}, while the result at $\sqrt{s} = 365$~GeV allows us  to put significant limits on $C_{eu}[1,1,3,3]$.

The right panel shows the bounds for the coefficients $C_{ u \phi}[3,3]$ and $C_{\phi u}[3,3]$. The $h \to gg$ channel at $\sqrt{s} =  240$~GeV constrains $C_{ u \phi}[3,3]$, as pointed out also in Fig.~\ref{fig:corr_240}, while $h \to b \bar{b}$ at $\sqrt{s} =  365$~GeV offers complementary information to derive the bounds on $C_{ \phi u}$. The limits on both sets of coefficients are greatly improved as compared to Fig.~3 of Ref.~\cite{Asteriadis:2024xuk}, which considers SMEFT NLO effects in the total rate for $e^+ e^- \to Zh$ only. 

Finally, in Fig.~\ref{fig:kappa} we see the sensitivity to $C_{\phi D}$ and $C_{\phi}$ on the left-hand side  and to $\delta_Z$ and $\kappa_\lambda$ on the right-hand side. 
We can reinterpret  our linear dimension-6 SMEFT calculation in terms of the parameters $\delta_Z$ and $\kappa_\lambda$ that are widely used \cite{McCullough:2013rea,Altmann:2025feg,Maura:2025rcv,Rossia:2023hen,terHoeve:2025omu} to describe new physics effects in the $hZZ/hWW$ couplings and the Higgs self-interaction, via the relations, \cite{Asteriadis:2024xts}
\begin{eqnarray}
\delta_Z &=& \frac{1}{4} \frac{v^2}{\Lambda^2} \biggl(C_{\phi D}+4 C_{\phi\square}\biggr) \nonumber \\
\delta \kappa_\lambda &=& \frac{v^2}{\Lambda^2} \left( 3\biggl[\frac{C_{\phi D}}{4}- C_{\phi \square}\biggr] - 2  \frac{v^2}{m_H^2} C_\phi \right) \, .
\end{eqnarray}

Including the Higgs decays at NLO, especially in the $h \to b \bar{b}$ and $h \to W W^*$ channels, is crucial for constraining the Higgs coupling to vector bosons and the Higgs tri-linear interaction and significantly improves the limits with respect to those from production alone, already in the $240$~GeV run, as seen on the left-hand side of Fig. \ref{fig:kappa}.
On the right-hand side, we combine the information from $\sqrt{s}=240$ and $365$ GeV obtained by considering only the production process (grey ellipses). Lighter colors correpond to $95\%$ C.L. and darker ones to $68\%$ C.L.
Without the data from the Higgs decays, it is necessary to run at two energies in order to obtain a constraint. The red circles in the right-hand panel of Fig. \ref{fig:kappa} corresponds to the information from $\sqrt{s}=240$ GeV including the decays, and we see a significant improvement over the grey ellipses with just the run at a single energy.
\section{Conclusions}
The understanding of Higgs interactions at a precise level is central to the goals of future colliders.  We have accomplished an important milestone in this program with the complete NLO EW plus QCD SMEFT dimension-6 calculation for all 2- and 3-body Higgs decays, computing for the first time the complete NLO EW results for $h\rightarrow WW^*$ and $h\rightarrow ZZ^*$ including all decay channels.   At the LHC, not all relevant Higgs production processes are known at NLO EW order for the SMEFT.  At future circular colliders, however, we have studied the process $e^+e^-\rightarrow Zh, \,  h\rightarrow$ all. Our results allow for a consistent NLO fit to production and decay and we find that including both production and decay significantly improves the limits on the SMEFT coefficients from those obtained using only the total rate for $e^+e^-\rightarrow Zh$.  In particular, the extraction at FCC-ee of $C_\phi$, which impacts the Higgs tri-linear coupling, is strengthened by including our results for the NLO SMEFT Higgs decays.    The impact of the Higgs decay information on the extraction of the Higgs tri-linear coupling was anticipated in the single parameter fit of \cite{Degrassi:2016wml}. Without the information from the Higgs decay, separating the effects of $C_\phi$ from other coefficients requires measurements at two different energies, while here we have explicitly shown how including the Higgs decays allows to significantly improve the sensitivity to $C_\phi$ in a single energy run. 

Both analytic and numerical results for the Higgs decays at NLO discussed here will be presented in a forthcoming paper, along with the flexible Monte Carlo code used to generate the plots in this work.
\section{Acknowledgments}
S. D. and C.D.P. are supported
by the U.S. Department of Energy under Contract No. DE-
SC0012704. P.P.G. is supported by the Ramón y Cajal grant~RYC2022-038517-I funded by MCIN/AEI/10.13039/501100011033 and by FSE+, and by the Spanish Research Agency (Agencia Estatal de Investigación) through the grant IFT Centro de Excelencia Severo Ochoa~No~CEX2020-001007-S. The work of L.B. is supported in part by the U.S.
Department of Energy under Grant No. DE-SC0010102 and by the College of Arts and Sciences of Florida State University. LB thanks the Technische Universität München (TUM) for the hospitality and the Excellence Cluster ORIGINS which is funded
by the Deutsche Forschungsgemeinschaft (DFG, German Research Foundation)
under Germany´s Excellence Strategy – EXC-2094 – 390783311 for
partial support during the completion of this work. The work of M.F. is supported by the U.S. National Science Foundation grant PHY-2210533.  Digital data is provided in the supplemental material attached to this paper. 

\newpage
\bibliographystyle{apsrev4-1}
\bibliography{refs.bib}
\end{document}